\documentclass[%
 reprint,
superscriptaddress,
 amsmath,amssymb,
 aps,
]{revtex4-2}

\usepackage{xcolor}
\usepackage{graphicx}% Include figure files
\usepackage{dcolumn}% Align table columns on decimal point
\usepackage{bm}% bold math
\usepackage{siunitx}

\usepackage{hyperref}% add hypertext capabilities
\hypersetup{
    colorlinks=true,
    linkcolor=blue,
    citecolor=blue,
    filecolor=blue,
    urlcolor=blue
}

\begin{document}

\preprint{APS/123-QED}

\title{Sawtooth wave adiabatic passage in a grating magneto-optical trap}

\author{Peter K. Elgee}
\email{peterelgee@gmail.com}
\affiliation{Joint Quantum Institute, University of Maryland and National Institute of Standards and Technology, College Park, MD}

\author{Ananya Sitaram}
\email{a.sitaram@uva.nl}
\affiliation{Joint Quantum Institute, University of Maryland and National Institute of Standards and Technology, College Park, MD}

\author{Sara Ahanchi}
\affiliation{Joint Quantum Institute, University of Maryland and National Institute of Standards and Technology, College Park, MD}

\author{Nikolai N. Klimov}
\affiliation{Sensor Science Division, National Institute of Standards and Technology, Gaithersburg, MD}

\author{Stephen P. Eckel}
\affiliation{Sensor Science Division, National Institute of Standards and Technology, Gaithersburg, MD}

\author{Gretchen K. Campbell}
\affiliation{Joint Quantum Institute, University of Maryland and National Institute of Standards and Technology, College Park, MD}

\author{Daniel S. Barker}
\email{daniel.barker@nist.gov}
\affiliation{Sensor Science Division, National Institute of Standards and Technology, Gaithersburg, MD}

\date{\today}

\begin{abstract}
We demonstrate sawtooth wave adiabatic passage (SWAP) in a grating magneto-optical trap (MOT) operating on the $^1$S$_0$\,$\rightarrow$\,$^3$P$_1$ transition of neutral $^{88}$Sr.
From numerical simulations of SWAP using our laser beam geometry, we find that SWAP provides greater cooling than triangle wave frequency modulation despite the complex polarization environment of a grating MOT.
The simulation is confirmed by our experimental results, where we demonstrate a factor of two improvement in transfer efficiency between our $^1$S$_0$\,$\rightarrow$\,$^1$P$_1$ grating MOT and our $^1$S$_0$\,$\rightarrow$\,$^3$P$_1$ grating MOT.
We trap up to $3\times10^6$ $^{88}$Sr atoms in the $^1$S$_0$\,$\rightarrow$\,$^3$P$_1$ grating MOT, at an average temperature of 4.9~$\mu$K with a lifetime of approximately 0.7~s.
Our results show that SWAP is effective in non-orthogonal laser beam geometries, allowing greater duty cycles or higher atom number in sensors based on narrow-line grating MOTs.
\end{abstract}

\maketitle

\section{\label{sec:Introduction}Introduction}

Laser-cooling has enabled the development of quantum technologies including quantum computers with trapped neutral atoms~\cite{Saffman10}, atomic clocks~\cite{Ludlow15}, and quantum sensors~\cite{Degen17}.
Laser-cooled alkaline-earth (AE) and AE-like atoms have attracted significant interest for these technologies due to their narrow, forbidden optical transitions.
In particular, AE atoms are employed for quantum sensors such as gravimeters~\cite{Hu2019}, atomic clocks~\cite{Bothwell2019, Hinkley2013}, gravitational wave detectors~\cite{Graham13,Vutha15}, and inertial navigation devices~\cite{Abend2023}.
For AE-atom-based sensors to operate outside laboratories -- either on satellites or at distributed ground stations -- they must be simple, stable, and small.

The narrow optical transitions that make AE atoms appealing as sensors also complicate the miniaturization of their laser-cooling systems.
As AE atoms lack the ground-state hyperfine structure needed for strong sub-Doppler cooling mechanisms, they typically require two sequential magneto-optical trap (MOT) cooling stages to reach the temperatures necessary for quantum sensors~\cite{Raab1987, Katori1999, Mukaiyama2003}.
Typically, a first-stage MOT operates on a broad electric-dipole-allowed transition with high capture velocity and a second-stage MOT operates on either a narrow electric-dipole-forbidden transition or a transition from a metastable state that permits sub-Doppler cooling~\cite{Katori1999, Grunert2002, Mukaiyama2003, Bowden19, Akatsuka2021}.
% For convenience, we will label the first MOT and second MOT as the broad-line MOT and narrow-line MOT, respectively.
Conventional MOTs require three orthogonal pairs of counter-propagating laser beams and have large optical layouts, which has motivated efforts to produce most or all of the laser beams for both the broad-line and narrow-line MOT using integrated optics~\cite{Rushton2014}.
Examples of compact MOT laser beam geometries that have been adopted for AE atoms include pyramid MOTs~\cite{Lee96, Pollock09, Bowden19, Pick24}, photonic-integrated-circuit MOTs~\cite{McGehee21, Isichenko23, Ropp23, Ferdinand2025}, Fresnel MOTs~\cite{Bondza24, Pick24}, and grating MOTs~\cite{Vangeleyn10, Nshii2013, Sitaram2020, Bondza2022}.
Recently, atom transfer from the broad-line MOT to a dipole-forbidden, narrow-line MOT has been demonstrated in the pyramid~\cite{Pick24}, photonic-integrated-circuit~\cite{Ferdinand2025}, Fresnel~\cite{Bondza24, Pick24}, and grating MOT geometries~\cite{Bondza2022}; allowing cooling to temperatures on the order of 10~$\mu$K.
The Fresnel and grating geometries are particularly advantageous because they combine high optical access with simple input laser beam routing~\cite{Vangeleyn10, Nshii2013, Bondza24}.

Fresnel and grating MOTs both employ non-orthogonal, tetrahedral laser beam geometries~\cite{Vangeleyn09, Vangeleyn10, Bondza24}.
In such geometries, the MOT laser beams do not have pure polarization when projected onto the local quantization axis defined by the MOT's quadrupole magnetic field~\cite{Lee13, Imhof2017}.
The impure polarization projection, combined with the efficiency of the diffraction grating or Fresnel reflector, leads to asymmetric cooling forces that are significantly reduced compared to a conventional MOT~\cite{Lee13, Imhof2017, Bondza2022, Barker2023}.
All compact MOT geometries also suffer from a small capture volume.
Because all quantum sensors benefit from high measurement duty cycle and large atom number, many approaches have been developed to increase the loading rate, trapped atom lifetime, and -- for AE atoms -- broad-line to narrow-line transfer efficiency of tetrahedral MOTs.
Examples include increasing the capture volume using 2D gratings~\cite{Nshii2013}, boosting the MOT loading rate with Zeeman slowers~\cite{Barker2019} or 2D MOTs~\cite{Imhof2017}, extending the trap lifetime with differential pumping~\cite{Imhof2017, Sitaram2020, Ehinger2022}, and improving atom transfer to the narrow-line MOT with achromatic reflectors~\cite{Bondza24, Pick24}.

Sawtooth wave adiabatic passage (SWAP) is another method to increase transfer efficiency between the broad-line and narrow-line MOTs~\cite{Muniz2018, Snigirev2019}.
The SWAP technique uses coherent transitions between the ground and excited states of a narrow optical transition with counter-propagating frequency-swept laser beams, allowing stimulated momentum transfers to enhance radiation pressure forces~\cite{Norcia2018, Bartolotta2018}.
Prior experimental studies of SWAP have occurred in one-dimensional, or six-beam, orthogonal laser beam geometries~\cite{Norcia2018, Muniz2018, Greve2019, Snigirev2019, Petersen2019} and prior simulations of SWAP have all used one-dimensional geometries with simplified atomic level structures~\cite{Bartolotta2018, Greve2019, Snigirev2019, Gan2020, Bartolotta2020}.
The tetrahedral laser beam geometry of grating and Fresnel MOTs potentially inhibits SWAP-enhancement of radiation pressure forces in three ways.
First, the non-orthogonal laser beams reduce the achievable momentum transfer per laser frequency sweep along any given axis.
Second, the impure projected polarization of the MOT laser beams reduces the amount of optical power addressing each excited Zeeman state, making the adiabaticity necessary for SWAP more difficult to achieve.
%Third, the distinct polarization projections of laser beams propagating parallel or oblique to the magnetic field promote shelving of atoms in the excited state, which causes the SWAP force to invert on sequential frequency sweeps~\cite{Norcia2018, Muniz2018}.
Third, the tetrahedral laser beam geometry promotes shelving of atoms in the excited state -- because there is no axis where any pair of laser beams both have a pure polarization projection -- which causes the SWAP force to invert on sequential frequency sweeps~\cite{Norcia2018, Muniz2018}.

Here, we study SWAP in the tetrahedral geometry of a grating MOT operating on the $^1$S$_0$\,$\rightarrow$\,$^3$P$_1$ transition of strontium.
In Sec.~\ref{sec:simulations}, we numerically simulate SWAP in our grating MOT geometry and find that it provides stronger optical forces than standard broadband MOT operation~\cite{Muniz2018, Snigirev2019}, despite the challenges outlined above.
We describe upgrades to the grating MOT apparatus of Ref.~\cite{Sitaram2020} in Sec.~\ref{sec:Apparatus}.
In Sec.~\ref{sec:SWAP}, we report our experimental realization of sawtooth wave adiabatic passage in a tetrahedral laser beam geometry and confirm the effectiveness of SWAP in improving atom transfer between our broad-line and narrow-line MOTs.
In Sec.~\ref{sec:NarrowLineMOT}, we measure the temperature and lifetime of our narrow-line grating MOT. % with up to $3\times10^6$ $^{88}$Sr atoms at an average temperature of approximately 4.9~$\mu$K and a lifetime of approximately 0.7~s.
We discuss our results and provide outlook for future experiments in Sec.~\ref{sec:outlook}.

\section{\label{sec:simulations}SWAP simulations}

During SWAP cooling, atoms are adiabatically transferred between the ground and excited states of a narrow-linewidth optical transition by counter-propagating frequency-swept laser beams.
% By sweeping the beams' frequency in a sawtooth pattern to coherently drive the transition many times, large amounts of energy can be removed from the system through stimulated forces without relying on spontaneous emission~\cite{Bartolotta2018, Norcia2018}.
Large amounts of energy can be removed from the system through stimulated forces without relying on spontaneous emission by sweeping the beams' frequency in a sawtooth pattern to coherently drive the transition many times~\cite{Bartolotta2018, Norcia2018}.
In a MOT, the addition of a magnetic field and the opposite polarizations of the counter-propogating beams causes the traditional mechanism for SWAP to break down~\cite{Muniz2018}.
However, even with some reliance on spontaneous emission, SWAP can still show enhancement over Doppler cooling in a strontium MOT~\cite{Muniz2018,Bartolotta2020}.

\begin{figure*}[t!]
  \center % chktex 1
  \includegraphics[width=2\columnwidth]{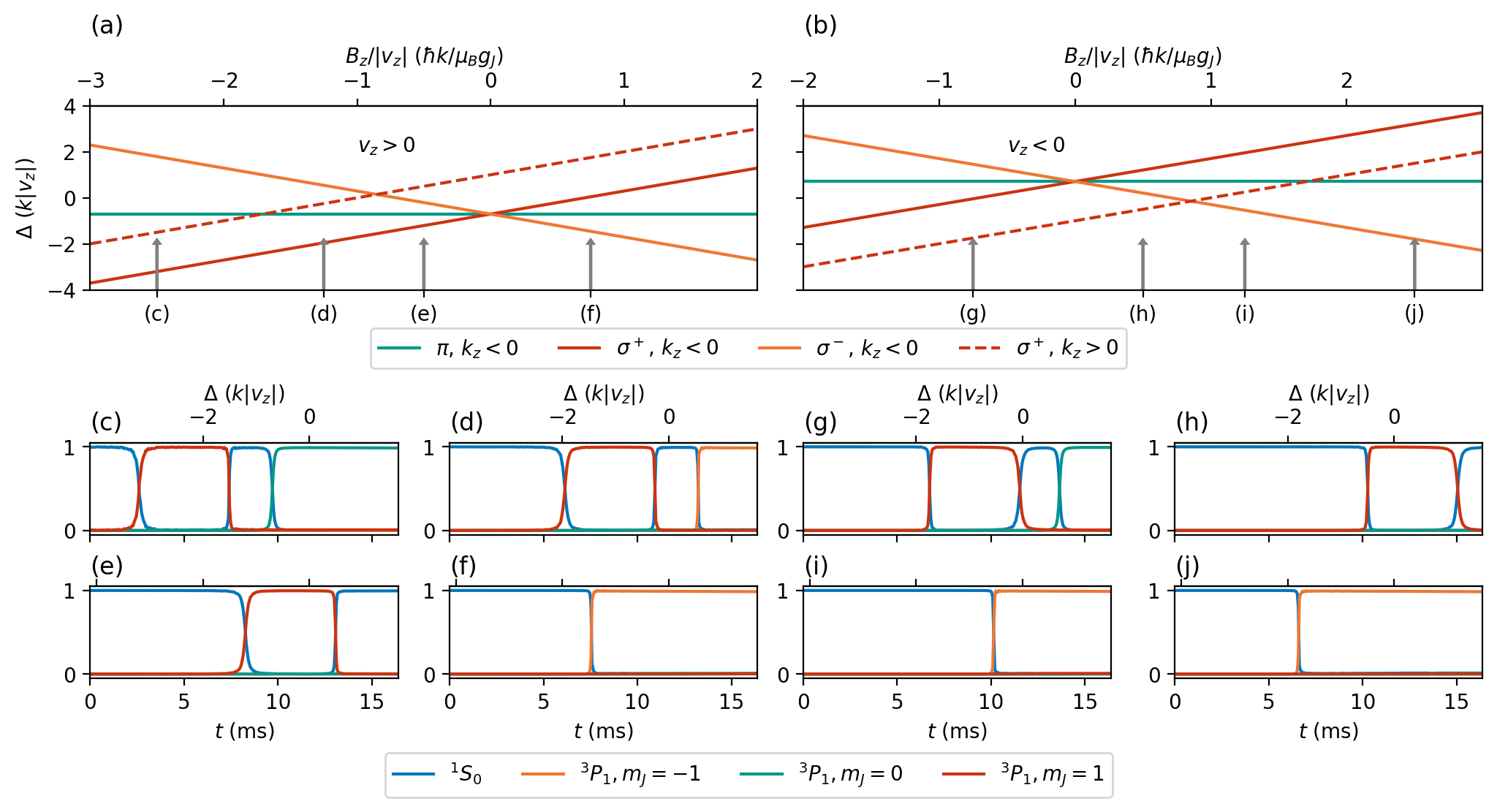}
  \caption{\label{fig:TrajFig}
  Possible transitions that can be made in a single laser frequency sweep.
  Subfigures (a) and (b) show the resonant laser detuning $\Delta$ of transitions excited by the input and diffracted laser beams as a function of the axial magnetic field $B_z$ normalized by the axial atomic speed $|v_z|$ for positive and negative velocity $v_z$, respectively.
  The dashed line indicates the transition due to the input beam and the solid lines indicate the transitions due to the diffracted beams.
  The arrows are labeled corresponding to the subfigures below, where the direction of the arrow indicates the direction of the frequency sweep across the transitions. 
  (c)-(j) show the populations of the atomic energy levels as a function of time and laser detuning for a single increasing sawtooth frequency sweep for an atom at the particular value of $B_z/|v_z|$ indicated by the corresponding arrow in subfigure (a) or (b).
  Subfigures (c)-(f) have positive $v_z$ and (g)-(j) have negative $v_z$.
  }
\end{figure*}

% In laser-cooling systems that use sequential MOTs, the efficiency of the transfer between the broad-line MOT and the narrow-line MOT is crucial to trap a sufficient number of atoms for experiments at a high repetition rate.
% However, the efficiency is limited by the mismatch in transition linewidth and the high temperature of the atoms in the broad-line MOT.
% Without large narrow-line cooling forces, the hot atoms in the broad-line MOT will escape the narrow-line MOT trapping region before they can be captured.
% Maximizing the narrow-line MOT cooling forces is of particular importance in a grating MOT geometry.
% Compared to traditional sequential six-beam MOTs, sequential grating MOTs typically have lower a initial atom number, a smaller trapping region, lower cooling forces, and incomplete overlap between the trapping regions of the two MOTs~\cite{Barker2023, Burrow23}.
% Sawtooth wave adiabatic passage significantly increases the cooling forces in the narrow-line MOT and is straightforward to implement in most sequential MOT systems.
% SWAP has been demonstrated to increase transfer efficiency into narrow-line, six-beam MOTs, but, to our knowledge, it has not been shown to increase transfer efficiency into a narrow-line, grating MOT.

Prior numerical studies have used a 1D model for SWAP and relied on the symmetry of the six-beam MOT geometry to make 3D predictions~\cite{Muniz2018, Snigirev2019}. 
Expanding the theory of SWAP to the non-orthogonal geometry of a grating MOT is complicated by the polarization and intensity gradients of the tetrahedral laser beam geometry~\cite{Lee13, Barker22, Barker2023}, and by off-axis atomic velocities causing additional Doppler shifts.
These complications increase the number of possible transitions per sawtooth period, and cause spatial variation in the adiabaticity requirements for these transitions.
Using the PyLCP python package~\cite{Eckel2022}, we implement a fully 3D optical Bloch equation (OBE) model of SWAP and gain understanding of how SWAP operates in the complex beam geometry of a grating MOT.

We explore the effect of the complex polarizations by describing the momentum transfer from SWAP along the axial direction $\hat{z}$, which is aligned with the input beam wavevector $\vec{k}$.
(The diffraction grating normal is $-\hat{z}$).
The polarization of the input beam is $\sigma^+$ and the polarization of the diffracted beams has $\sigma^+$, $\pi$, and $\sigma^-$ components, in decreasing optical power order (in a traditional MOT, the counterpropagating beam has only $\sigma^-$ character)~\cite{Raab1987, McGilligan15, Imhof2017, Barker2019, Bondza2022}.
As a result, transitions to and from any excited state Zeeman state are possible.
Whether, and when, a transition to a particular Zeeman state occurs in a given laser frequency sweep is dependent on the current state of the atom and the ratio of the Zeeman shift of the state to the Doppler shift of the beam stimulating the transition.

To explore all possible adiabatic passages, we neglect the effects of spontaneous emission by setting the natural decay rate to an artificially low value of $\Gamma=1$~s$^{-1}$.
We adjust the sweep time $t_s$ -- which is the duration of the laser frequency sweep from initial detuning $\delta_i$ to final detuning $\delta_f$ -- and the Rabi frequency $\Omega$ of the input beam such that every possible transition is adiabatic (\(\Omega^2t_s/|\delta_f-\delta_i|\gg1\)).
Figure~\ref{fig:TrajFig} shows all possible adiabatic transfers as the laser detuning is swept over all resonances for atoms moving with velocity $v_z$ along $\hat{z}$ as a function of $B_z$, which is the magnetic field projection along $\hat{z}$ (and $\hat{z}$ points towards the grating chip).
The resonant laser detuning $\delta_r$ of each transition is shown in Fig.~\ref{fig:TrajFig}(a) and Fig.~\ref{fig:TrajFig}(b) for positive and negative $v_z$, respectively.
Figure~\ref{fig:TrajFig}(c) to Fig.~\ref{fig:TrajFig}(j) show the adiabatic transfers that occur when an atom is at the distinct $B_z/|v_z|$ indicated by the arrows in Fig.~\ref{fig:TrajFig}(a) and Fig.~\ref{fig:TrajFig}(b) during the frequency sweep.
For simplicity, the atomic motion along the transverse $x$ and $y$ axes is frozen in the simulations shown in Fig.~\ref{fig:TrajFig}(c) to Fig.~\ref{fig:TrajFig}(j).

There are 8 distinct SWAP regimes as a function of $v_z$ and $B_z$, with different momentum transfer $\Delta p_z$ per sweep outlined in Tab.~\ref{tab:momentum transfers} and indicated in Fig.~\ref{fig:TrajFig}.
Both the boundary between regimes and $\Delta p_z$ in each regime depend on the grating diffraction angle $\theta_d$.
In some regimes, the atom ends the sweep in an excited state (see Fig.~\ref{fig:TrajFig}), so spontaneous emission is necessary to transfer the atom back to the ground state and achieve cooling~\cite{Muniz2018, Bartolotta2020}.
As indicated in Tab.~\ref{tab:momentum transfers}, certain regimes exhibit a decrease in momentum $\Delta |p_z|$ greater than the expected momentum transfer of $-2\hbar k$ per sweep (for 1D SWAP cooling) and $-\hbar k$ per sweep (for a SWAP-enhanced, six-beam MOT), where $\hbar$ is the reduced Planck constant~\cite{Muniz2018,Norcia2018, Bartolotta2018,Snigirev2019}.
The cooling force depends on the time ordering of the transitions and, consequently, on sweeping the optical frequency from below to above the bare transition.
If the sweep direction is reversed -- a condition we refer to as anti-SWAP -- then the time ordering of transitions is also reversed, causing most of the adiabatic transfers to heat the atoms.
The simulations in Fig.~\ref{fig:TrajFig} provide intuition, but are not entirely realistic; in addition to the artificial natural decay rate and Rabi frequencies, they do not include interference of the diffracted beams or allow for transverse velocities.

\begin{table*}[t]%The best place to locate the table environment is directly after its first reference in text
\caption{\label{tab:momentum transfers}%
Momentum transfers for different regimes in velocity and magnetic field.}
\begin{ruledtabular}
\begin{tabular}{cccc}
 Fig.~\ref{fig:TrajFig} subfigure&sign $v_z$&$B_z/|v_z|$ range $(\hbar k / \mu_b g_J)$&$\Delta |p_z|$ ($\hbar k$)\\
\colrule
  (c) &+&$\qquad \qquad \ \ B_z/|v_z|<-1-\cos(\theta_d)$ &$-1 - 2\cos(\theta_d)$\\
  
  (d) &+& $-1-\cos(\theta_d)< B_z/|v_z| < -(1+\cos(\theta_d))/2$   & $-1 - 2\cos(\theta_d)$\\
  
  (e) &+&$-(1+\cos(\theta_d))/2< B_z/|v_z| < 0\qquad\qquad\qquad\qquad\:$  & $-1 - \cos(\theta_d)$\\
  
  (f) &+&$0 < B_z/|v_z|\qquad\quad\ \ $  & $-\cos(\theta_d)$\\
  
  (g) &$-$&$B_z/|v_z|<0$ & $-1$\\
  
  (h) &$-$&$\qquad \quad \: \: \ 0< B_z/|v_z| < (1+\cos(\theta_d))/2$  & $-1 - \cos(\theta_d)$\\
  
  (i) &$-$&$(1+\cos(\theta_d))/2< B_z/|v_z| < 1+\cos(\theta_d) \qquad \quad \: \: \: $  & $\cos(\theta_d)$\\
  
  (j) &$-$&$1+\cos(\theta_d)<B_z/|v_z|\qquad \qquad \qquad \quad \:\ $ & $\cos(\theta_d)$\\
\end{tabular}
\end{ruledtabular}
\end{table*}

To understand whether SWAP can enhance transfer into a narrow-line grating MOT, we simulate atomic trajectories with initial conditions suggested by the regimes of Fig.~\ref{fig:TrajFig}, including accurate spontaneous emission, transverse atomic motion, laser interference, and experimentally achievable Rabi frequencies.
Adding spontaneous emission causes the populations in Fig.~\ref{fig:TrajFig} to decohere.
Allowing transverse velocities splits the transitions from the different diffracted beams due to Doppler shifts, yielding more possible transitions per sweep. %, and in a real system many of these transitions will not be resolved in the $\fix{window}$ window required for smooth adiabatic transfer.
Additionally, atoms move through an interference pattern created by the diffracted beams that causes spatial variation in the adiabaticity of each transition.
%Lastly, the different powers in each polarization from the diffracted beams will have different adiabaticity requirements, which may not all be satisfied at once in a given system.
%The partial transfers from these power concerns can sometimes be aided by the additional diffracted beams which will couple to the remaining ground state population.
%If we model the system with realistic experimental parameters we see that the transfers become much more complex, but on average the advantage of SWAP holds.
% To understand whether SWAP can enhance transfer into a narrow-line grating MOT, we simulated atomic trajectories with initial conditions suggested by the regimes of Fig.~\ref{fig:TrajFig} including spontaneous emission.
% \fix{list parameters}
% \fix{
% \begin{itemize}
%     \item sweep frequency 20~kHz
%     \item decay rate $2\pi*7.5$~kHz
%     \item rabi 2e6 MHz
%     \item low det sawtooth = -2e6 Hz
%     \item high det sawtooth = 2e6 Hz
%     \item low det triangle -4e6 Hz
%     \item high det triangle 0
%     \item vels = 20e5, 15e5, 10e5, 5e5 / 2pi doppler shift Hz
%     \item b = -0.5*20e5 / 2 pi zeeman shift? Hz
% \end{itemize}
% }
We compare the atomic trajectories for sawtooth and triangle frequency sweeps using the natural decay rate $\Gamma \approx 2\pi\times7.5\times10^3$~s$^{-1}$ for the $^3$P$_1$ state, and an input beam Rabi frequency of $\Omega = 2\pi\times2$~MHz.
The sawtooth wave has $t_s=50$~$\mu$s and the triangle wave has $t_s=25$~$\mu$s, which gives both waveforms the same period (the period of the sawtooth waveform is $t_s$ and the period of the triangle waveform is $2t_s$).
The sawtooth sweep range starts at detuning $\delta_i = -2\pi\times2$~MHz and ends at detuning $\delta_f = 2\pi\times2$~MHz.
The triangle sweep range starts at $\delta_i = -2\pi\times4$~MHz and ends at $\delta_f = 2\pi\times0$~MHz to avoid blue detuning.

\begin{figure}[t!]
  \center % chktex 1
  \includegraphics[width=\columnwidth]{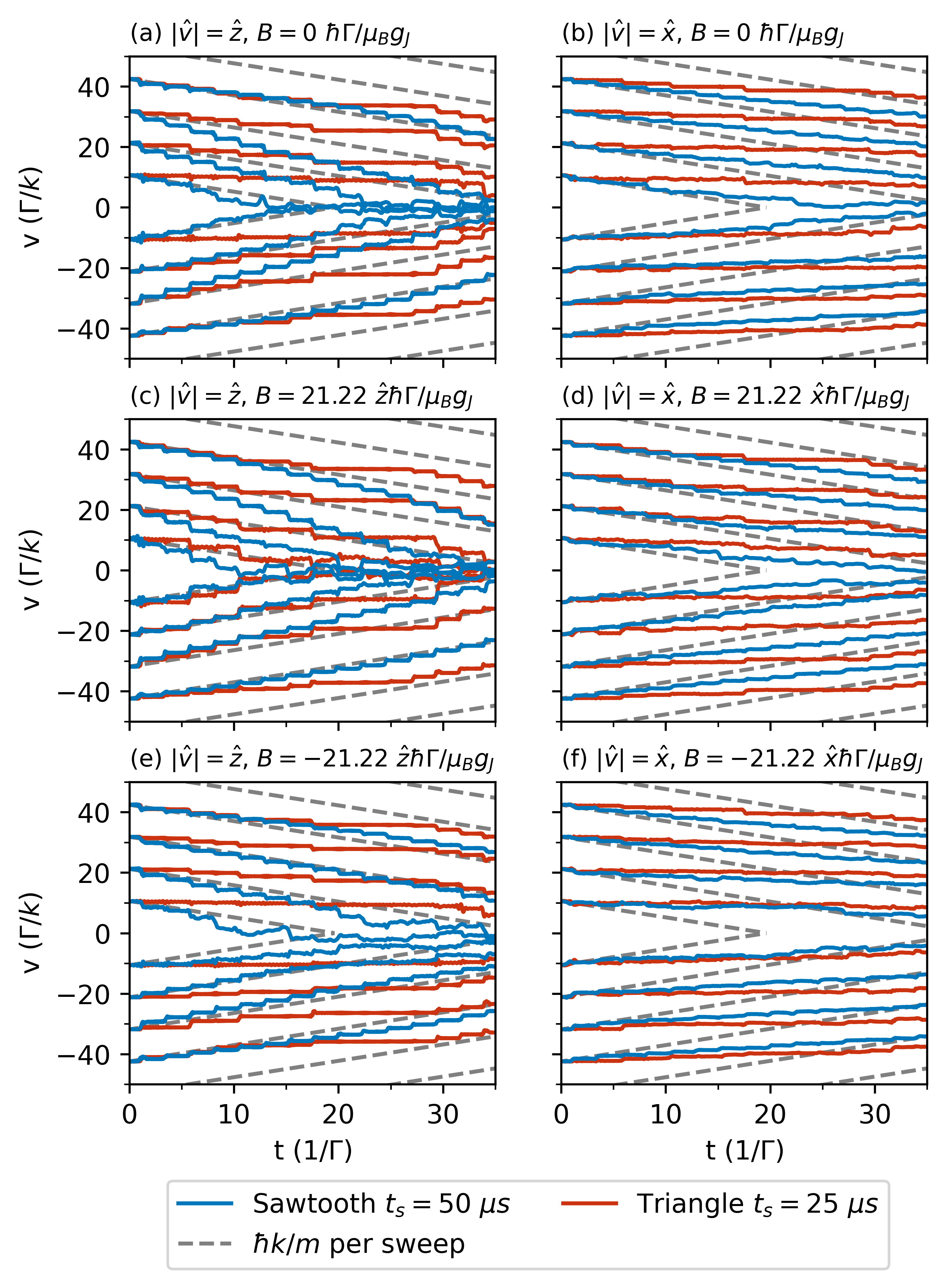}
  \caption{\label{fig:VelocityTraces}
  Examples of the atomic velocity over time for a variety of parameters.
  Subplots (a), (c), and (e) show axial velocities (along $\hat{z}$).
  Subplots (b), (d), and (f) show transverse velocities (along $\hat{x}$).
  Subplots (a) and (b) use zero magnetic field, while (c), (d), (e) and (f) use a non-zero field along $\hat{z}$ or $\hat{x}$ (as indicated near the subplot label).
  The red traces show the atomic velocities for a triangular frequency sweep, while the blue traces show the atomic velocities for a sawtooth frequency sweep (SWAP).
  The gray dashed lines represent a change in velocity of $\hbar k / m$ per sweep, which is the maximum cooling expected from SWAP in a six-beam MOT.
  }
\end{figure}

We compare sawtooth modulation with traditional triangle modulation in Fig.~\ref{fig:VelocityTraces}.
We simulate the effect of both sweep shapes for positive and negative velocities along $\hat{z}$ and $\hat{x}$ that correspond to Doppler shifts of $\pm2$~Mrad/s, $\pm1.5$~Mrad/s, $\pm1$~Mrad/s and $\pm0.5$~Mrad/s.
In addition, we simulate at zero magnetic field, and at magnetic fields that correspond to a Zeeman shift magnitude of 1~Mrad/s for the magnetically sensitive Zeeman states.
The simulations include the interference of the diffracted beams and are averaged over the initial position of the atom to account for the optical interference effect.
For the parameters explored in Fig.~\ref{fig:VelocityTraces}, SWAP cools more efficiently than a triangle sweep, and can provide velocity damping larger than the six-beam MOT limit of $\hbar k/m$ per sweep for initial velocities along $\hat{z}$, where $m$ is the atomic mass.
The simulations show that SWAP should enhance the transfer between the broad-line and narrow-line grating MOTs for parameters that are achievable in our apparatus.

\section{\label{sec:Apparatus}Apparatus}

\begin{figure}[t!]
  \center % chktex 1
  \includegraphics[width=\columnwidth]{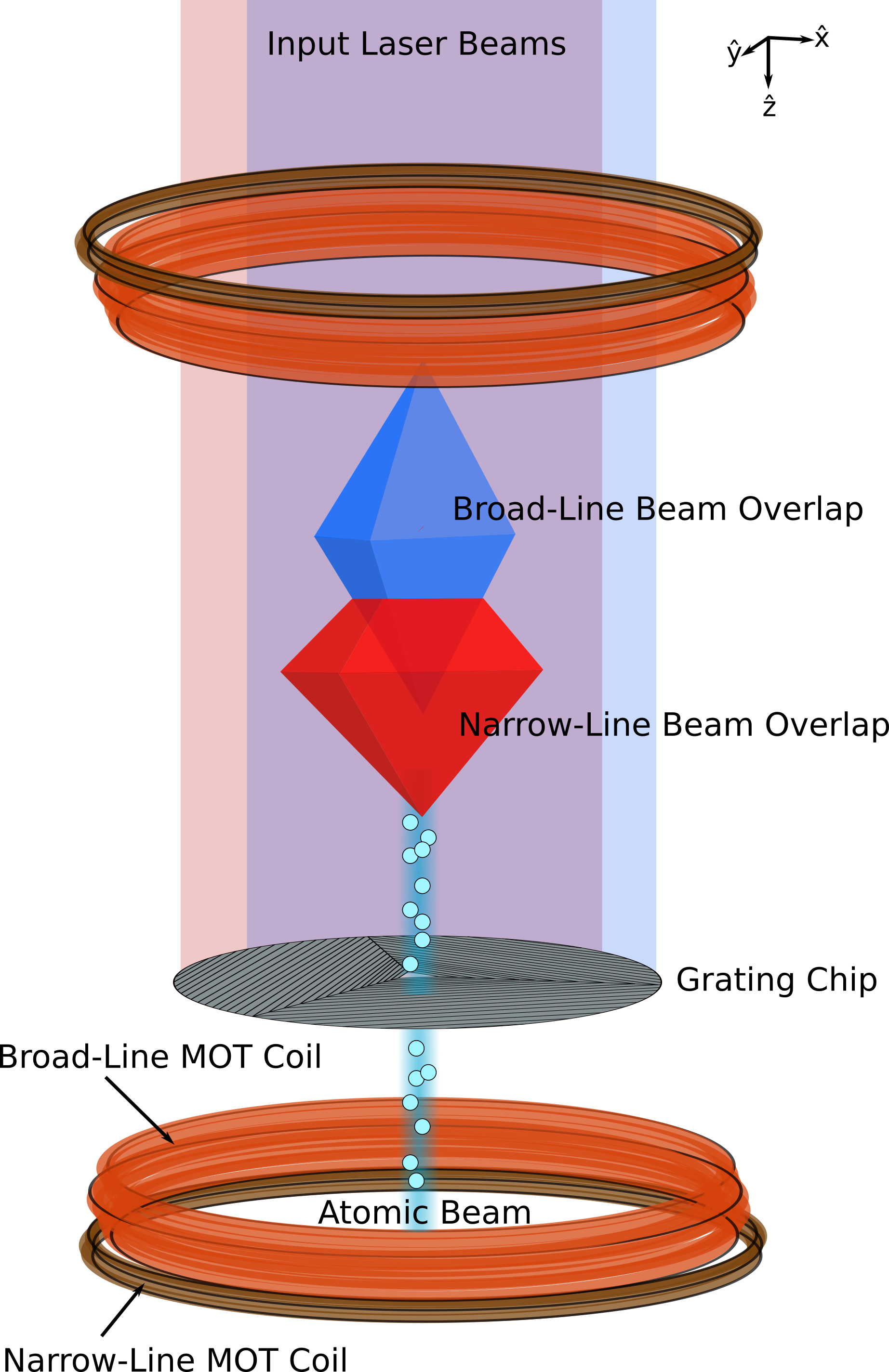}
  \caption{\label{fig:apparatus}
  Schematic of experimental apparatus.
  The atomic beam (light blue line) propagates in the $-\hat{z}$ direction (against gravity) and enters the broad-line MOT beam overlap through a triangular hole in the diffraction grating chip.
  The broad-line MOT and narrow-line MOT beam overlap volumes indicate the region where all four MOT laser beams have non-zero intensity for each MOT.
  Two sets of electromagnets create the broad-line MOT magnetic field gradient (orange) and narrow-line MOT magnetic field gradient (brown).
  The input laser beams for both MOTs enter the vacuum chamber from the top and propagate in the $+\hat{z}$ direction.
  }
\end{figure}

Our apparatus, diagrammed in Fig.~\ref{fig:apparatus}, has previously been described in Ref.~\cite{Sitaram2020}.
Here, we limit our discussion to upgrades made to achieve narrow-line cooling.
First, we replaced the permanent magnets used in Ref.~\cite{Sitaram2020} with electromagnets.
The new apparatus has two pairs of electromagnets aligned vertically (along $\hat{z}$, see Fig.~\ref{fig:apparatus}) in a quadrupole configuration.
The larger pair of electromagnets (labeled as the broad-line MOT coil in Fig.~\ref{fig:apparatus}) provides an approximately $5.5$~mT/cm gradient for the broad-line MOT and can be switched off in less than 1~ms.
The smaller pair of electromagnets (labeled as the narrow-line MOT coil in Fig.~\ref{fig:apparatus}) is mounted around the larger quadrupole electromagnets and provides a magnetic field gradient up to $0.4$~mT/cm for the narrow-line MOT.
The top and bottom electromagnet of this pair are controlled independently with bipolar current controllers, which allows them to provide a uniform vertical magnetic field in addition to the narrow-line MOT magnetic field gradient.
Lastly, two pairs of electromagnets aligned to the $\hat{x}$ and $\hat{y}$ directions (not shown in Fig.~\ref{fig:apparatus}) provide transverse magnetic fields to adjust the position of the narrow-line MOT's quadrupole field zero.

Second, the aluminum mounting block for the grating chip was replaced with a titanium mounting block to avoid eddy currents generated close to the atomic cloud when the magnetic field switches between the broad-line and narrow-line MOT field gradients.
Elimination of eddy currents is particularly important in a narrow-line grating MOT as the magnetic field gradient is small and the atoms are close to the mounting block.
After changing the mounting block, the vacuum pressure was improved to less than $1\times10^{-7}$~Pa.

Third, a 689~nm narrow-line MOT input beam was added using a dichroic mirror to combine it with the 461~nm broad-line MOT input beam \cite{Elgee22, Sitaram22}.
To make optical access for the narrow-line MOT input beam, the repump beams for the broad-line MOT now enter the vacuum chamber through a side viewport.
The narrow-line MOT input beam has a $1/e^2$ radius of approximately 17~mm, with a maximum peak intensity of $I_p\approx18$~mW/cm$^2$ (corresponding to $I_p\approx6000\, I_{\rm sat}$, where $I_{\rm sat}$ is the saturation intensity of the $^1$S$_0$\,$\rightarrow$\,$^3$P$_1$ transition).
The grating diffraction angle is $27.0(5)^\circ$ at 461~nm and $42.7(9)^\circ$ at 689~nm (here, and throughout the paper, parenthetical quantities indicate the standard uncertainty).
As a result, the trapping regions for the two MOTs are vertically offset with approximately 43~\% of the narrow-line MOT trapping region lying within the broad-line MOT trapping region (see Figure~\ref{fig:apparatus}).

\section{\label{sec:SWAP}Demonstration of SWAP cooling}

To demonstrate SWAP cooling, we begin by preparing a broad-line MOT containing approximately $3\times10^7$ $^{88}$Sr atoms, which we transfer to the narrow-line MOT through a two stage process.
First, we lower the power of the broad-line input laser beam by approximately a factor of 10 for 10~ms.
Second, the broad-line input laser beam and the large electromagnets are switched off to release the atoms into the narrow-line MOT.
We take the time at which the broad-line MOT is turned off as $t=0$~s.
The narrow-line input laser beam is on throughout the broad-line MOT and its frequency is swept using an acousto-optic modulator in a sawtooth pattern with $t_s = 50$~$\mu$s over an initial detuning range spanning from $\delta_i(0)=-2\pi\times6$~MHz to $\delta_f(0)=2\pi\times0$~MHz.
We then reduce the frequency sweep range over $t_{\rm ramp}=243$~ms to $\delta_i(t_{\rm ramp})=\delta_f(t_{\rm ramp})=-2\pi\times0.7$~MHz.
While ramping the frequency sweep range, we reduce the peak intensity of the narrow-line cooling beam exponentially from $I_p(0)=18$~mW/cm$^2$ to $I_p(t_{\rm ramp})=2$~mW/cm$^2$, which corresponds to $I_p(t_{\rm ramp})\approx670\,I_{\rm sat}$.
The small electromagnets for the narrow-line MOT are set to a gradient of $0.4$~mT/cm throughout the experiment.
The parameters above correspond to optimal transfer efficiency between the two MOTs.

\begin{figure}[t!]
  \center % chktex 1
  \includegraphics[width=\columnwidth]{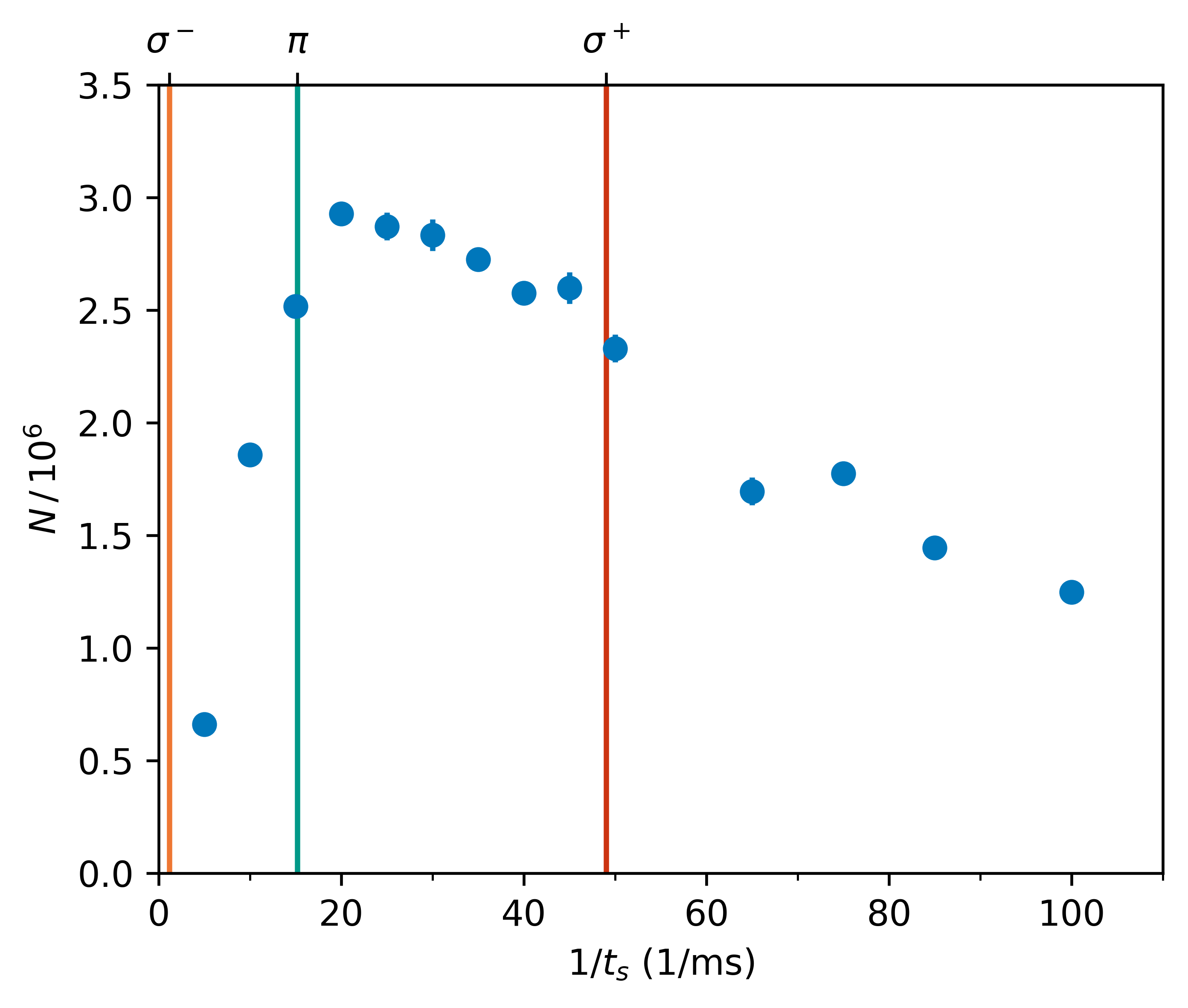}
  \caption{\label{fig:SWAPFrequency} Transferred atom number $N$ as a function of $1/t_s$ for SWAP.
  The vertical lines show the sweep time where adiabaticity breaks down (given by $\Omega^2 t_s/|\delta_f(0)-\delta_i(0)| = 1$) for each polarization component in a single diffracted beam at the start of the transfer.}
\end{figure}

We investigate the dependence of the atom number transferred to the narrow-line MOT $N$ as a function of $t_s$ in Fig.~\ref{fig:SWAPFrequency}.
Figure~\ref{fig:SWAPFrequency} also shows the sweep times at which adiabaticity breaks down (given by $\Omega^2 t_s/|\delta_f(0)-\delta_i(0)| = 1$) for each polarization component of a single diffracted laser beam at the start of the transfer sequence~\cite{Norcia2018}.
The peak in $N$ occurs close to the breakdown of adiabaticity for the $\pi$ polarized component and before the breakdown of adiabaticity for the $\sigma^+$ polarization component.
Adiabaticity breakdown for the input laser beam would occur at $t_s=5.65$~$\mu$s ($1/t_s=177$~ms$^{-1}$). %177~kHz.
The optimal $t_s=50$~$\mu$s yields approximately $3\times10^6$ atoms.
The data in Fig.~\ref{fig:SWAPFrequency} indicate that, at its optimal transfer efficiency, the narrow-line MOT does not rely on adiabatic $\sigma^-$ transitions from the diffracted beams.

\begin{figure}[t!]
  \center % chktex 1
  \includegraphics[width=\columnwidth]{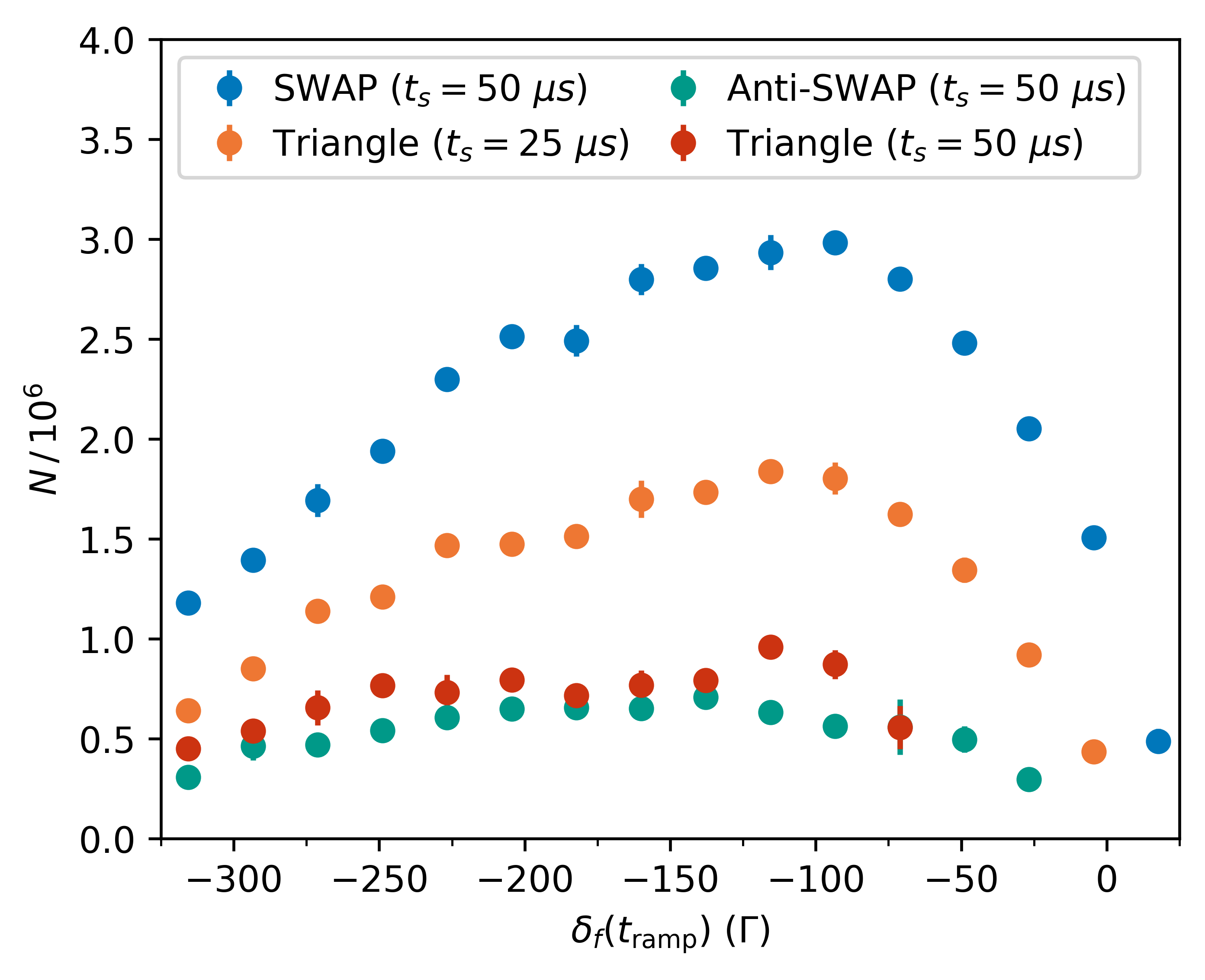}
  \caption{\label{fig:SWAPDetuning}Transferred atom number $N$ as a function of final detuning at the end of the ramp sequence $\delta_f(t_{\rm ramp})$ for different sweep times and shapes.
  The blue, orange, green, and red circles show data for SWAP sweep at $t_s=50$~$\mu$s, triangle sweep at $t_s=25$~$\mu$s, anti-SWAP sweep at $t_s=50$~$\mu$s, and triangle sweep at $t_s=50$~$\mu$s, respectively. 
  }
\end{figure}

We ran our broad-line to narrow-line MOT transfer with $t_s=50$~$\mu$s while scanning the detuning to compare the transferred atom number for a SWAP sweep, triangle sweep, and anti-SWAP sweep (sawtooth frequency sweep with negative slope, see Sec.~\ref{sec:simulations}).
For the triangle sweep, we also tested the transfer with $t_s=25$~$\mu$s, which is the sweep time that optimizes the six-beam narrow-line MOT on our other apparatus \cite{Elgee22, Sitaram22}.
In our detuning scans, we vary $\delta_i(0)$, $\delta_f(0)$, $\delta_i(t_{\rm ramp})$, and $\delta_f(t_{\rm ramp})$ with $\delta_f(0)-\delta_i(0)$, $\delta_f(t_{\rm ramp})-\delta_i(t_{\rm ramp})$, and $\delta_i(0)-\delta_i(t_{\rm ramp})$ all fixed (\textit{i.e.}, we vary the overall detuning of our detuning sweep).
Figure~\ref{fig:SWAPDetuning} shows $N$ in our narrow-line MOT as a function of $\delta_f(t_{\rm ramp})$ for each sweep shape.
The data for the triangle sweep with $t_s=50$~$\mu$s were taken with a lower broad-line MOT atom number and were scaled to match the data for the other sweeps.

We see that SWAP enhances $N$ by approximately a factor of two compared to a triangle sweep with $t_s=25$~$\mu$s.
% , and a factor of three compared to a triangle wave with $t_s=50$~$\mu$s.
The enhancement is larger than that found from SWAP in the six-beam $^{88}$Sr MOT presented in Ref.~\cite{Muniz2018}.
The larger enhancement could arise because of larger cooling forces in the grating MOT geometry (due to the possibility of more than two momentum transfers per sweep, see Sec.~\ref{sec:simulations}), or a greater importance of cooling efficiency in our system (due to the small narrow-line MOT beam overlap volume, see Sec.~\ref{sec:Apparatus}).
The triangle wave with $t_s=50$~$\mu$s provides lower atom number than the triangle wave with $t_s=25$~$\mu$s.
Anti-SWAP yields the lowest atom number, as expected given the importance of time ordering in SWAP: when the sweep direction is reversed, most atoms are heated rather than cooled. 

\section{\label{sec:NarrowLineMOT}Narrow-line MOT performance}

\begin{figure}[t!]
  \center % chktex 1
  \includegraphics[width=\columnwidth]{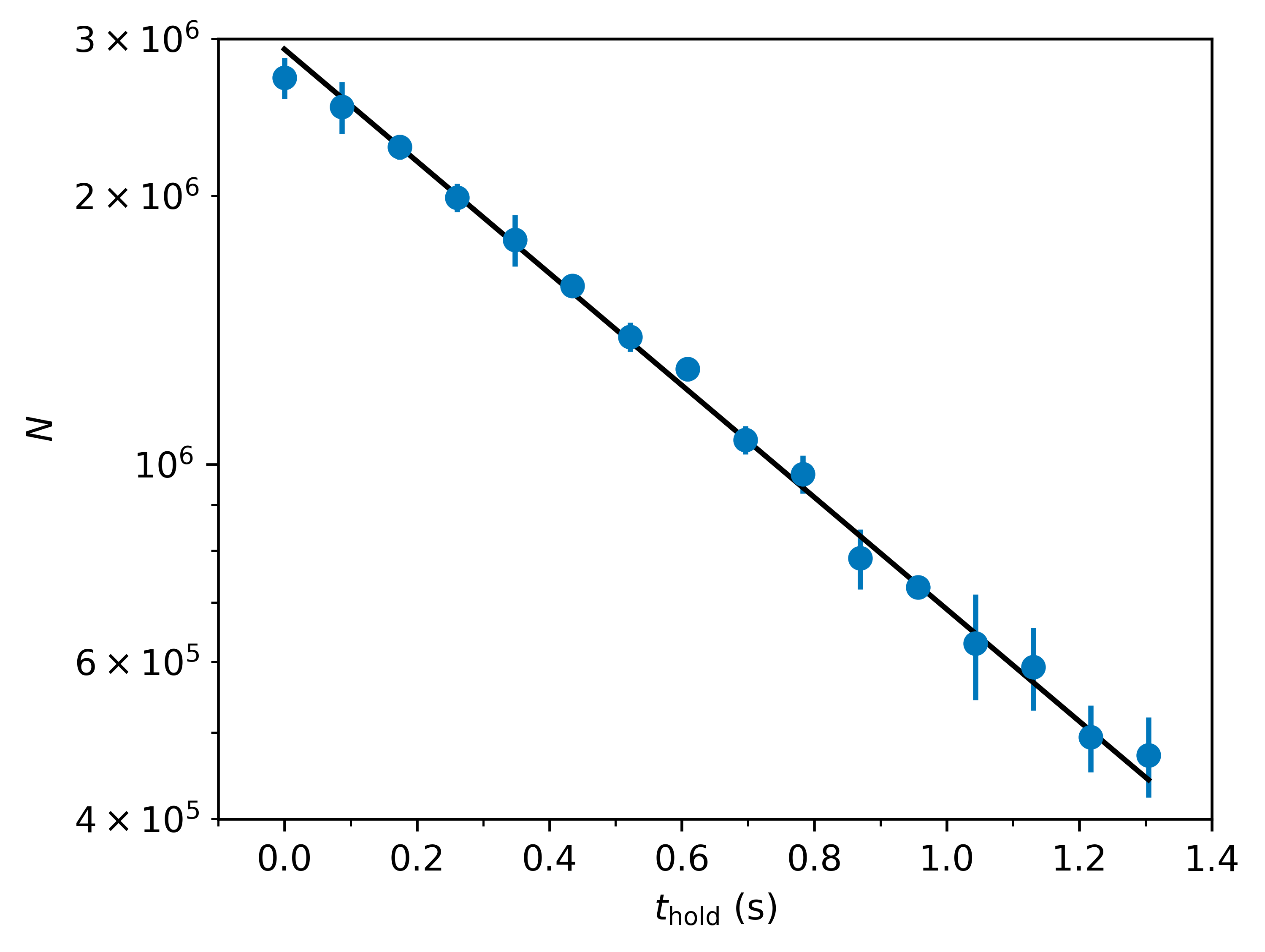}
  \caption{\label{fig:Lifetime}
  Atom number $N$ as a function of MOT hold time $t_{\rm hold}$ for the narrow-line MOT.
  The solid black line is an exponential fit, which yields a lifetime of approximately 0.7~s.
  }
\end{figure}

\begin{figure}[t!]
  \center % chktex 1
  \includegraphics[width=\columnwidth]{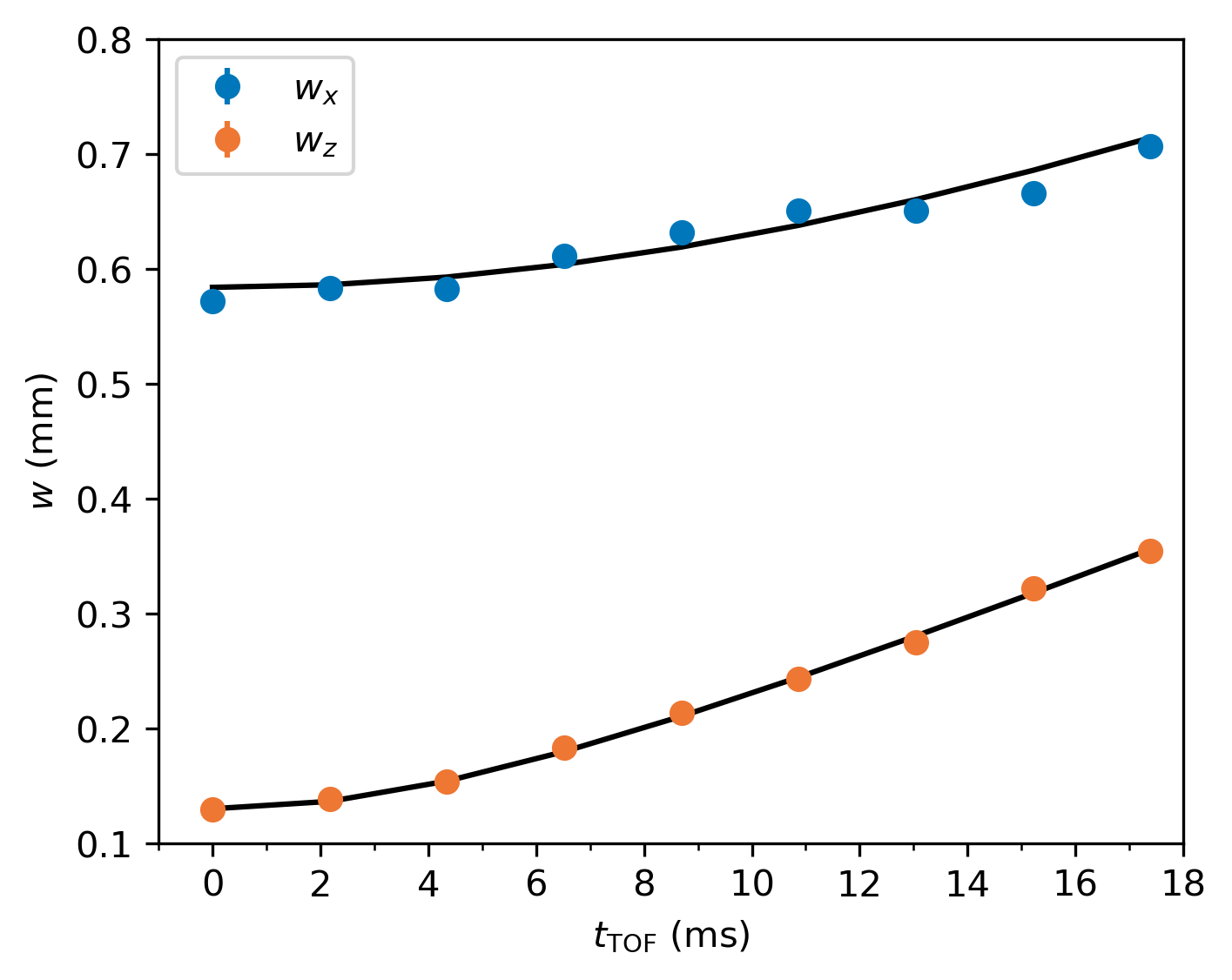}
  \caption{\label{fig:Temperature}
Temperature measurement for the narrow-line $^{88}$Sr grating MOT. 
The root-mean-square width of the atomic cloud in the transverse (blue circles) and axial (orange circles) directions are plotted as a function of time of flight.
The data are fit to the expansion curve described in the text (black lines), giving a transverse temperature of 5.9(6) $\mu$K and an axial temperature of 3.8(1) $\mu$K.
The error bars are smaller than the data points and represent the standard error in the mean.}
\end{figure}

We characterize the lifetime and temperature of the narrow-line MOT using the optimized SWAP transfer discussed in Sec.~\ref{sec:SWAP}.
Fig.~\ref{fig:Lifetime} shows the decay of the atom number in the narrow-line MOT with hold time $t_{\rm hold}$ after the frequency sweep ramp.
An exponential fit to the decay data yields a $1/e$ lifetime of approximately 0.7~s.
We measured the axial and transverse temperatures using the root-mean-square width of the cloud in time of flight, shown in Fig.~\ref{fig:Temperature}.
We fit the widths to ${w(t_{\rm TOF})}^2 = w_{0}^2 + v^2_{rms}t_{\rm TOF}^2$, where $t_{\rm TOF}$ is the time of flight, $w_0$ is the initial width of the cloud, $v_{rms}=\sqrt{k_BT/m}$ is the root-mean-square velocity, $k_B$ is Boltzmann's constant, $m$ is the atomic mass, and $T$ is the temperature of the atomic cloud.
The axial temperature is $3.8(1)~\mu$K, and the transverse temperature is $5.9(8)~\mu$K.
The temperature anisotropy is expected due to the geometry of the grating MOT~\cite{McGilligan15}.
The average temperature of approximately $5~\mu$K is higher than is typically achieved in six-beam narrow-line strontium MOTs~\cite{Katori1999, Loftus2004}, but similar to those reported for narrow-line grating or Fresnel MOTs~\cite{Bondza2022, Bondza24, Pick24}.

Despite our realization of a broad-line grating MOT of $^{87}$Sr~\cite{Barker2023}, we have not been able to trap $^{87}$Sr in the narrow-line grating MOT.
We believe this is due to spin polarization from the input beam negating the effect of the conventional `stirring' scheme that stabilizes six-beam $^{87}$Sr MOTs~\cite{Mukaiyama2003, Barker2023}.
In addition to conventional stirring, we attempted to stabilize the narrow-line $^{87}$Sr grating MOT using SWAP (without stirring as suggested by Ref.~\cite{Muniz2018}) and an unconventional stirring scheme using the $F = 9/2 \rightarrow F' = 7/2$ transition (as suggested by Ref.~\cite{Barker2023}), but were not successful.
Finding a stable narrow-line grating MOT configuration for $^{87}$Sr may require a different grating diffraction angle and efficiency~\cite{Barker2023, Burrow23}. 

\section{\label{sec:outlook}Outlook}

We have investigated and demonstrated SWAP cooling in the tetrahedral laser beam geometry of a grating MOT.
Optical Bloch equation simulations of SWAP show significant shelving of atoms in the excited state due to the impure polarization projection of diffracted beams onto the local quantization axis in our laser beam geometry.
As a result, spontaneous emission plays a more important role in generating cooling than in 1D counter-propagating laser beam arrangements.
Nevertheless, OBE simulations of atomic trajectories in the tetrahedral laser beam geometry indicate that SWAP provides faster cooling than traditional triangle wave frequency modulation.
Experimentally, we observe that SWAP improves atom transfer from our broad-line grating MOT to our narrow-line grating MOT by a factor of two.
Using SWAP, we can trap as many as $3\times10^6$ $^{88}$Sr atoms in our narrow-line grating MOT at an average temperature of approximately $5$~$\mu$K.
Our results show that SWAP can be an effective tool to increase atom number or measurement duty cycle in miniaturized laser-cooled-atom-based sensors.
% The complex polarizations yield more than two possible adiabatic transfers per sweep, potentially providing more cooling force than in a traditional six-beam MOT along the principle axes of the field gradient, while the variable powers in each beam and polarization complicate the adiabaticity of these transfers.
% Additionally, while the SWAP utilizes adiabatic transfers to provide cooling forces beyond what is achievable through spontaneous emission alone, in both a traditional and grating MOT the technique still relies on spontaneous emission to ensure the atoms are preferentially initialized in the ground state at the start of a sweep. 
% The enhancement of cooling forces from SWAP compliment the particular difficulties of the grating MOT where cooling forces are generally lower, and the volume of the trapping region is small.
% We demonstrate this enhancement through a factor of two increase in our atom number when using SWAP.
% Our analysis and demonstration of SWAP in a grating MOT shows how we can achieve more efficient loading in compact cold atom systems, in particular our narrow-line grating MOT of $^{88}$Sr, and have taken a step closer to field deployable quantum devices using alkaline-earth atoms.
% Our validation of SWAP in a strontium grating MOT addresses transfer efficiency issues in the platform and further supports its potential as a method for transitioning quantum technologies from the laboratory to the field.

\begin{acknowledgments}

We thank Z. Ahmed and M. Hummon for their careful reading of the manuscript.
Our work is funded by the National Institute of Standards and Technology.

\end{acknowledgments}

% \appendix

% \section{Appendixes}

\bibliography{SWAPMOT}% Produces the bibliography via BibTeX.

\end{document}